# Development of a GPU-accelerated Monte Carlo dose calculation module for nuclear medicine, ARCHER-NM: Demonstration for a PET/CT imaging procedure

Zhao Peng[1, 2], Yu Lu[1, 2], Yao Xu[1, 2], Yongzhe Li[1, 2], Bo Cheng[1, 2], Ming Ni[3], Zhi Chen[1, 2], Xi Pei[1, 2], Qiang Xie[3], Shicun Wang[2,3], X. George Xu[1, 2, 4]

[1] School of Nuclear Science and Technology, University of Science and Technology of China, Hefei 230026, China

[2] Institute of Nuclear Medical Physics, University of Science and Technology of China, Hefei 230026, China

[3] Department of Nuclear Medicine, The First Affiliated Hospital of USTC, Division of Life Science and Medicine, University of Science and Technology of China, Hefei 230001, China

[4] Department of Radiation Oncology, The First Affiliated Hospital of USTC, University of Science and Technology of China, Hefei 230001, China

*Corresponding author:*

*Professor X. George Xu, Ph.D., F.AAPM, F.AIMBE, F.HPS, F.ANS*

*University of Science and Technology of China*

*Huangshan Road 443, Hefei 230026, Anhui, China*

*Email: xgxu@ustc.edu.cn*

**Abstract**

This paper describes the development and validation of a Monte Carlo (MC) dose computing module dedicated to organ dose calculations of patients undergoing nuclear medicine (NM) internal radiation exposures involving $^{18}$F-FDG PET/CT examination. This new module extends the more-than-10-years-long ARCHER project that developed a GPU-accelerated MC dose engine by adding dedicated NM source-definition features. To validate the code, we compared dose distributions from the 0.511-MeV point photon source calculated for a water phantom as well as a patient PET/CT phantom against a well-tested MC code, GATE. The water-phantom results show excellent agreement, suggesting that the radiation physics module in the new NM code is adequate. To demonstrate the clinical utility and advantage of ARCHER-NM, one set of PET/CT data for an adult male NM patient is calculated using the new code. Radiosensitive organs in the CT dataset are segmented using a CNN-based tool called DeepViewer. The PET image intensity maps are converted to radioactivity distributions to allow for MC radiation transport dose calculations at the voxel level. The dose rate maps and corresponding statistical uncertainties were calculated for the duration of PET image acquisition. The dose rate results of the $^{18}$F-FDG PET imaging patient show that ARCHER-NM's results agree very well with those of the GATE within 0.58% to 4.11% (for a total of 27 organs considered in this study). Most impressively, ARCHER-NM obtains such results in less than 0.5 minutes while it takes GATE as much as 376 minutes for the same number of $10^9$ simulated decay events. This is the first study presenting GPU-accelerated patient-specific MC internal radiation dose rate calculations for clinically realistic $^{18}$F-FDG PET/CT imaging case involving autosegmentation of whole-body PET/CT images. This study suggests that modern computing tools —— ARCHER-NM and DeepViewer —— are accurate and fast enough for routine internal dosimetry in NM clinics.

## 1. Introduction

Both positron emission tomography (PET) and computed tomography (CT) are widely used imaging modalities in the diagnosis and monitoring of cancer evolution. The combination of PET and CT, known as PET/CT, enables both anatomical and metabolic imaging of patients, which improves the diagnostic quality and efficiency of radiologists (Kapoor *et al.*, 2004). However, in PET/CT examination procedures, the injection of radiopharmaceuticals results in internal ionizing radiation for patients, which increases the risk of radiation-induced cancer, particularly in younger patients (Huang *et al.*, 2009; Belinato *et al.*, 2017). To quantify such risks, it is critical to calculate the absorbed doses of the patients with enough accuracy and speed acceptable for routine clinical applications (Einstein *et al.*, 2007). Therefore, it is essential to develop a method for internal dosimetry in this field.

Traditionally, the internal dosimetry of a specific patient can be estimated at the organ level via the S-value —— the mean absorbed dose in a target organ per radioactivity decay in a source organ calculated in a standard phantom (Loevinger *et al.*, 1988; Stabin and Siegel, 2003). Based on this method, Andersson al et. (Andersson *et al.*, 2017) developed an internal dosimetry program IDAC-Dose 2.1 to estimate the absorbed dose for diagnostic nuclear medicine using specific absorbed fraction values of the ICRP computational voxel phantoms. But this method assumes homogeneous activity and dose distributions in organs and a generalized geometry; hence, it does not consider patient-specific activity distributions and organ anatomies (Gupta *et al.*, 2019). More precise methods to address the internal dosimetry at the voxel level have been developed, including the convolution of dose point kernels (Giap *et al.*, 1995) and the voxel S-value approach (Bolch *et al.*, 1999; Bolch *et al.*, 2009; Amato *et al.*, 2012; Amato *et al.*, 2013a; Amato *et al.*, 2013b). However, these methods usually assume a uniform human tissue material density, neglecting the differences between the lung tissue, soft tissue, and bone (Amato *et al.*, 2013a; Moghadam *et al.*, 2016). Furthermore, the point-kernel approach is truncated to a limited range near the calculation point to trade accuracy for efficiency. (Auditore *et al.*, 2019). These limitations can lead to certain inaccuracies in the dose calculation.

On the other hand, Monte Carlo (MC) methods have long been a computing tool for internal dose distributions at the voxel level. Direct MC simulations coupled with functional and anatomical imaging are considered the gold standards for patient-specific dose estimation (Zaidi, 1999; Zaidi and Xu, 2007; Neira *et al.*, 2020). Several general-purpose MC codes can be used to calculate the radiation dose, including Geant4 (Agostinelli *et al.*, 2003), MCNP (Yoriyaz *et al.*, 2001), and FLUKA (Botta *et al.*, 2013). Geant4 is one of the most validated and widely used, notably for medical physics purposes (Allison *et al.*, 2016). Geant4 Application for Tomography Emission) (GATE) (Jan *et al.*, 2004) is a toolkit that offers a user-friendly interface for Geant4, is specific for emission tomography, and is suited for dose calculations in radiotherapy and nuclear medicine (NM) applications (Sarrut *et al.*, 2014). GATE has been validated fully and used widely in recent years (Parach *et al.*, 2011; Jan *et al.*, 2011; Hickson and O'Keefe, 2014; Villoing *et al.*, 2017; Gupta *et al.*, 2019; Pistone *et al.*, 2020). However, MC simulations are notoriously slow due to the massive amount of calculations required to reach an acceptable accuracy, which prevents their clinical use.

In the last decade, the use of general-purpose graphics processing units (GPUs) for Monte Carlo radiation transport simulations has emerged (Hissoiny *et al.*, 2011; Jia *et al.*, 2011; Jia *et al.*, 2015; Jia *et al.*, 2014; Pratx and Xing, 2011), bringing impressive parallel computational efficiency to the method that was thought unfit for clinical workflow. In our work on ARCHER (Su *et al.*, 2014), a dedicated GPU-based MC code has been validated for a wide range of medical physics applications, such as CT imaging and radiotherapy (Xu *et al.*, 2015; Adam *et al.*, 2020; Lin *et al.*, 2017). In this work, an ARCHER's capabilities are extended by adding a new NM module dedicated to internal dosimetry for patients undergoing PET/CT examination involving $^{18}$F-FDG PET/CT imaging. To this end, the NM module is designed to perform rapid individualized MC dose calculations on the CT image data of the patient, with radioactivity distributions constructed from the PET image data. At the same time, the key radiosensitive organs of the whole body are segmented by an autosegmentation technique. Finally, the organ dose rates are obtained, since PET images are acquired at a specific time. The overall computational flow is shown in Fig. 1. To the best of our knowledge, this is the first study presenting GPU-accelerated patient-specific MC internal radiation dose rate calculations and autosegmentation of whole-body image dataset for a clinically realistic $^{18}$F-FDG PET/CT imaging case. The comparison of organ dose rates of ARCHER-NM against GATE MC simulations is performed using identical $^{18}$F-FDG PET/CT data. This work aimed to validate and benchmark the ARCHER-NM module for internal dosimetry of PET imaging.

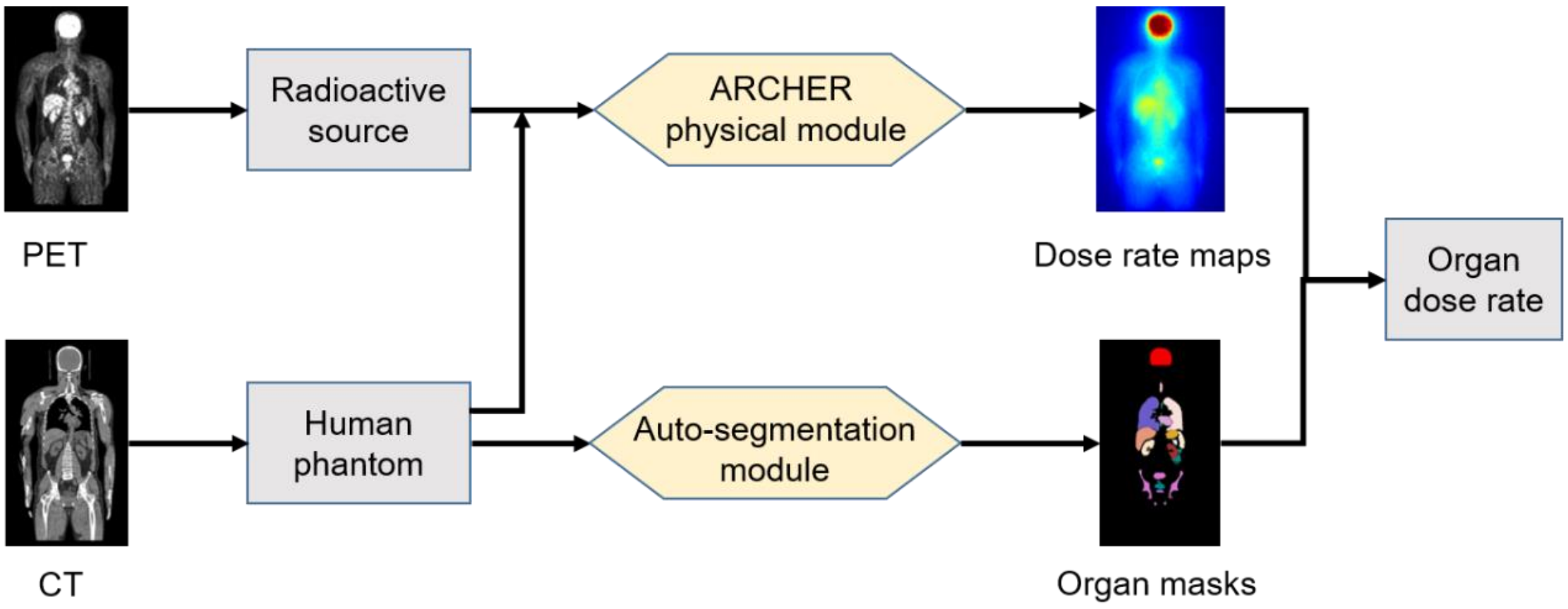

Fig. 1. The overall flow of ARCHER-NM for calculating the organ dose rate in PET imaging.

## 2. Methods

### 2.1. ARCHER-NM setup

For internal radiation dose rate calculations involving the PET/CT imaging patient, ARCHER-NM is designed to use CT images to construct a human phantom. The CT images are converted into mass density and material composition using the Hounsfield unit (HU)-to-density conversion curve [39]. In this process, four materials, each having a density specified by the HU, are used for the patient phantom: water, dry air, compact bone (defined by ICRU), and lung (defined by ICPR) (Schneider *et al.*, 2000; Kawrakow and Walters, 2006). The PET images indicate the intensity distribution of radionuclides expressed in Bq/mL (Pistone *et al.*, 2020). The probability density distribution of the decay events in MC simulations is constructed through a linear conversion of the PET value. Each nuclear decay leads to a positron that undergoes annihilation, resulting in two photons of equal energy (0.511 MeV) in opposite directions. The isotropic angular distribution is defined using Marsaglia method to sample uniformly from the surface of a sphere (Marsaglia, 1972).

In ARCHER-NM, the photoelectric effects and Compton scattering are explicitly considered, while the pair production is ignored since the photon energy is below the threshold of 1.02 MeV. A photon or an electron is transported in the human CT phantom until its energy falls below cutoff energy. At this point, the MC transport simulation is terminated, and its remaining energy is deposited locally. The cutoff energies for photons and electrons are set to 10 keV and 200 keV, respectively. The deposited dose is calculated for each voxel, and the corresponding uncertainty is evaluated using the history-by-history method (Walters *et al.*, 2002; Chetty *et al.*, 2006). The dose and uncertainty maps have the same size and resolution as those of the input human phantom.

The obtained 3D absorbed dose maps are divided by the total number of decay events in the simulation to deduce dose-per-event maps. Multiplying each dose-per-event voxel by the whole body total radioactivity at the PET scan acquisition time ($t = t_s$), we calculate dose rate maps $\dot{D}(t_s)$ (Gy/s) at the acquisition time $t_s$ according to (Pistone *et al.*, 2020):

$$\dot{D}(t_s) = \frac{D}{N_{evts}} \times A(t_s) \quad (1)$$

where $D$ (Gy) is the total absorbed dose, $N_{evts}$ is the total number of decay events in the MC simulations, and $A(t_s)$ (Bq) is the whole body total activity measured in the PET images at the acquisition time $t_s$.

### 2.2. Simulation setup

First, we perform a simulation experiment of a point photon source in water. Here, we create CT images and PET images with a size of 101×101×101 and a voxel spacing of 0.5 mm × 0.5 mm × 0.5 mm. All voxels in CT images are assigned a value of 0 to simulate the water box. All the voxels in PET images are assigned a value of 0 except for the voxel at the center which is assigned a value of 1 as a point photon source. The number of simulated decay events is $1 \times 10^8$ for both the GATE and ARCHER-NM.

Second, we perform internal radiation dose rate calculations for a whole-body $^{18}$F-FDG PET imaging patient. One set of PET/CT data for an adult male is collected from the Department of Nuclear Medicine of the First Affiliated Hospital of the University of Science and Technology of China (Hefei, China). This patient is 52 years old, 170 cm in height, and 79 kg in weight. The amount of injected $^{18}$F-FDG is 8.0 mCi. Whole-body PET scanning is performed after a 60-minute waiting period when the patient has urinated. The size of PET images is 168×168×219, and the voxel spacing is 4.063 mm × 4.063 mm × 5 mm. The CT images are resampled to obtain the same size as the PET images. The number of simulated decay events is $1 \times 10^9$ for both GATE and ARCHER-NM.

GATE (version 9.0) of Geant4 (10.6.2) is used in this study. To improve the calculational speed of GATE, 25 threads are used for each simulation experiment of GATE. The CPU is an Intel® Xeon® Gold 5120T @ 2.20 GHz. ARCHER-NM simulations are executed using an NVIDIA Titan V GPU.

### 2.3. Organ autosegmentation

To evaluate the organ dose of patients undergoing PET/CT examination, corresponding organs must be segmented. However, it would be impractical to manually delineate so many organs from the image dataset. Herein, we used the deep-learning-based organ autosegmentation software DeepViewer (Wisdom Tech, Hefei, China; http://www.wisdom-tech.com.cn/) to solve this problem. The accuracy of segmentation has been validated in our previous studies (Peng *et al.*, 2020; Wang *et al.*, 2020). Specifically, the Dice Similarity Coefficients (DSC) is above 0.9 for most organs such as the brain, lung, heart, liver, and so on. In this study, based on the CT images, 27 key organs of the whole body were autosegmented. To ensure the accuracy of the segmentation, the segmentation results were checked by a clinical doctor.

### 2.4. Evaluation standard

For the dose distribution calculation of a point photon source in water, the dose-distance curves are compared between GATE and ARCHER-NM. For the dose rate calculation of $^{18}$F-FDG PET imaging patients, the average dose rate $\dot{d}(t_s)$ and corresponding uncertainties ε in the organs are calculated according to (Chetty *et al.*, 2006):

$$\dot{d}(t_s) = \frac{1}{N}\sum_{i=1}^{N} \dot{D}_i\,(t_s) \tag{2}$$

$$\varepsilon = \sqrt{\frac{1}{N}\sum_{i=1}^{N} \varepsilon_i^2} \tag{3}$$

where $N$ is the number of voxels in an organ and $\dot{D}_i(t_s)$ and $\varepsilon_i$ are the dose rate and uncertainty of voxel $i$ in the acquisition time $t_s$, respectively. Taking GATE as a reference, the relative percent differences $\sigma$ are calculated according to:

$$\sigma = 100 \times \frac{\dot{d}(t_s)_{ARCHER-NM} - \dot{d}(t_s)_{GATE}}{\dot{d}(t_s)_{GATE}} \tag{4}$$

The dose rate maps of sagittal and coronal slices for GATE and ARCHER-NM are also compared. In addition, the simulation times for GATE and ARCHER-NM are analyzed.

## 3. Results

### 3.1. Point photon source

Fig. 2 shows the dose-distance curves for the internal radiation calculation of the point photon source in water. Here, the red lines with dots represent the results of GATE, and the blue lines with crosses represent the results of ARCHER-NM. In the dose region above 10% of the maximum dose, the relative percent difference of the average dose is less than 0.3% relative to results from GATE. These are excellent dose results from ARCHER-NM, suggesting the radiological physics aspects of the new module are modeled correctly. In terms of computing efficiency, ARCHER-NM is found to take only 2.4 seconds to run the MC simulations with a statistical uncertainty of 0.47%, while GATE takes approximately 1296 seconds for the same number of particles with a statistical uncertainty of 0.46%. In other words, while the dose results are practically identical, ARCHER-NM is roughly 540 times faster than GATE for the chosen experiment.

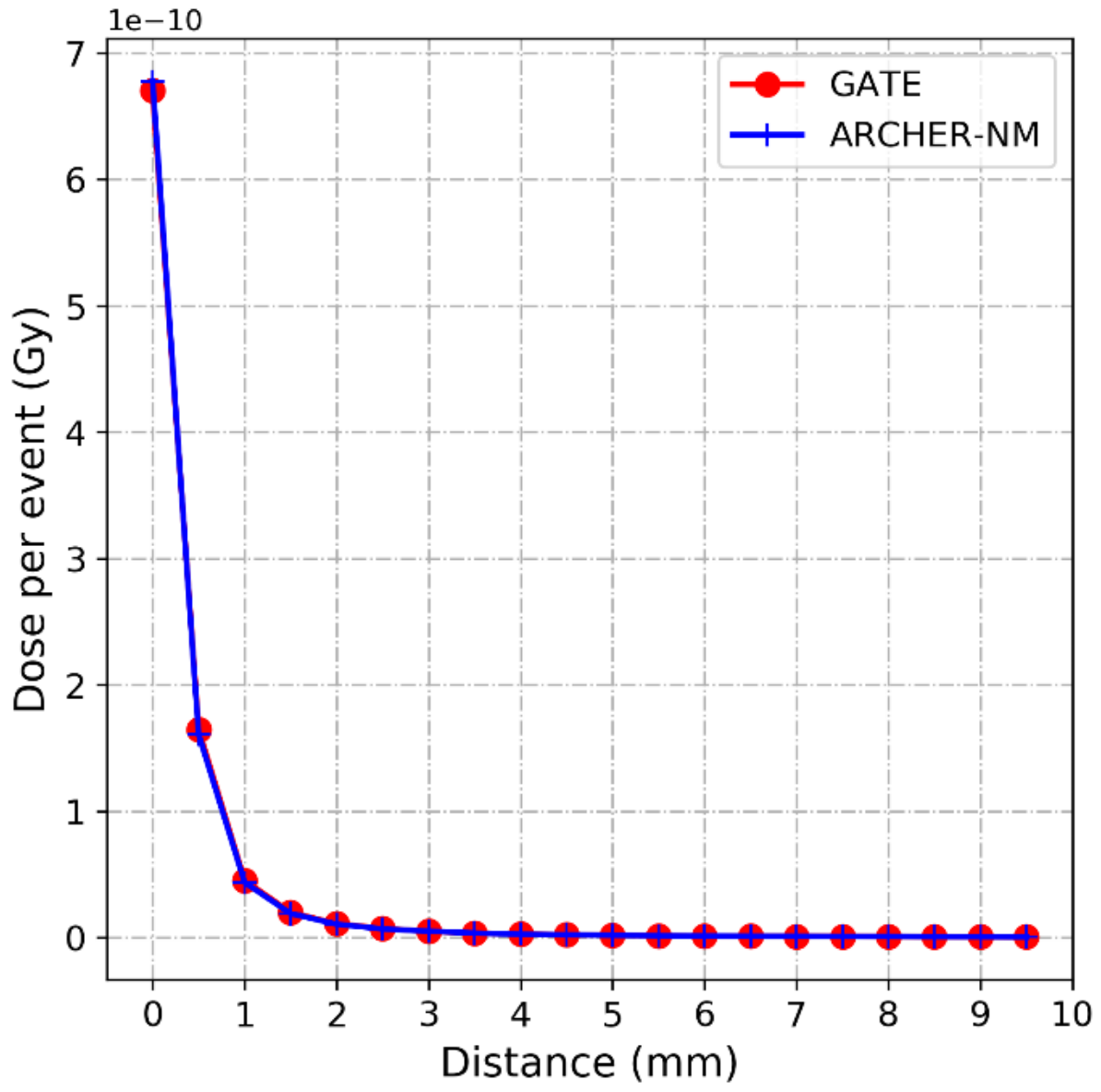

Fig. 2 Dose-distance relationship of each decay event for a 0.511-MeV point photon source in water showing excellent agreement between ARCHER-NM and GATE. The voxel size of the dose matrix is 0.5 mm $\times$ 0.5 mm $\times$ 0.5 mm.

### 3.2. $^{18}$F-FDG PET imaging

For the dose rate calculation involving the $^{18}$F-FDG PET imaging patient, it takes GATE 376 minutes to yield a statistical uncertainty of 0.99%, while ARCHER-NM takes 0.5 minutes to yield a statistical uncertainty of 1.19%. ARCHER-NM is 750 times faster than GATE for simulations involving the whole body of the patient anatomy. Fig. 3 shows the dose rate maps of sagittal and coronal views for the $^{18}F$-FDG PET imaging patient. The results of GATE and ARCHER-NM are displayed in the first column and second column, respectively. The relative difference maps are displayed in the third column, taking the GATE results as a reference. Here, the dose rates in the air are set to zero. The results indicate that the dose rate distributions are nearly identical between GATE and ARCHER-NM. The high-dose-rate areas appear to localize in the brain, as expected.

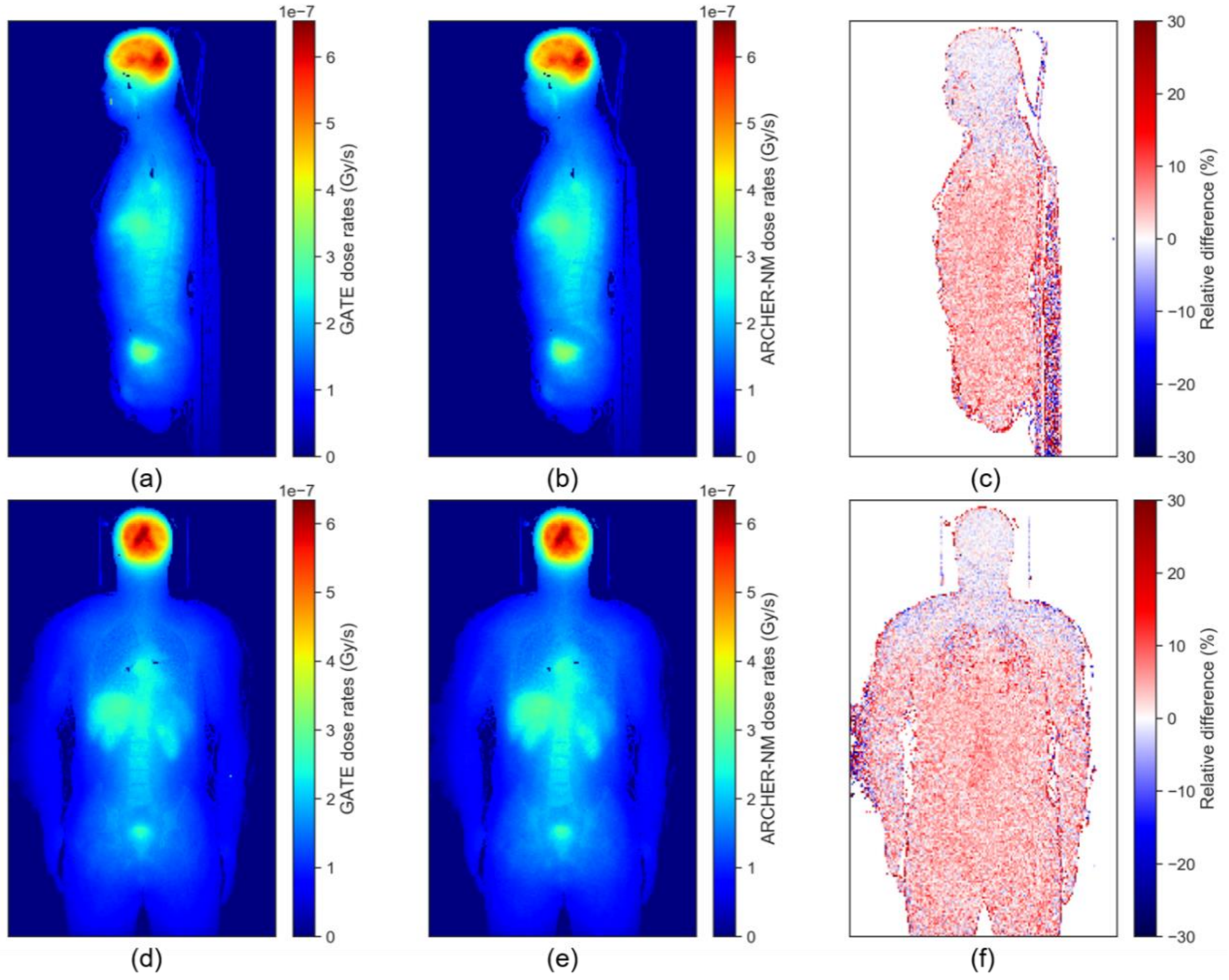

Fig. 3. Comparison of dose rate maps in sagittal and coronal views between results from GATE (a, d) and ARCHER-NM (b, e) for the $^{18}F$-FDG PET imaging patient. Relative error maps are displayed (c, f) taking GATE results as reference.

The autosegmentation results of a total of 27 organs are visualized in Fig. 4, in terms of sagittal and coronal 2D views as well as the 3D views. The total organ segmentation time was approximately 12 minutes consisting of an autosegmentation time of 7 minutes and a radiation oncologist's check-up time of 5 minutes.

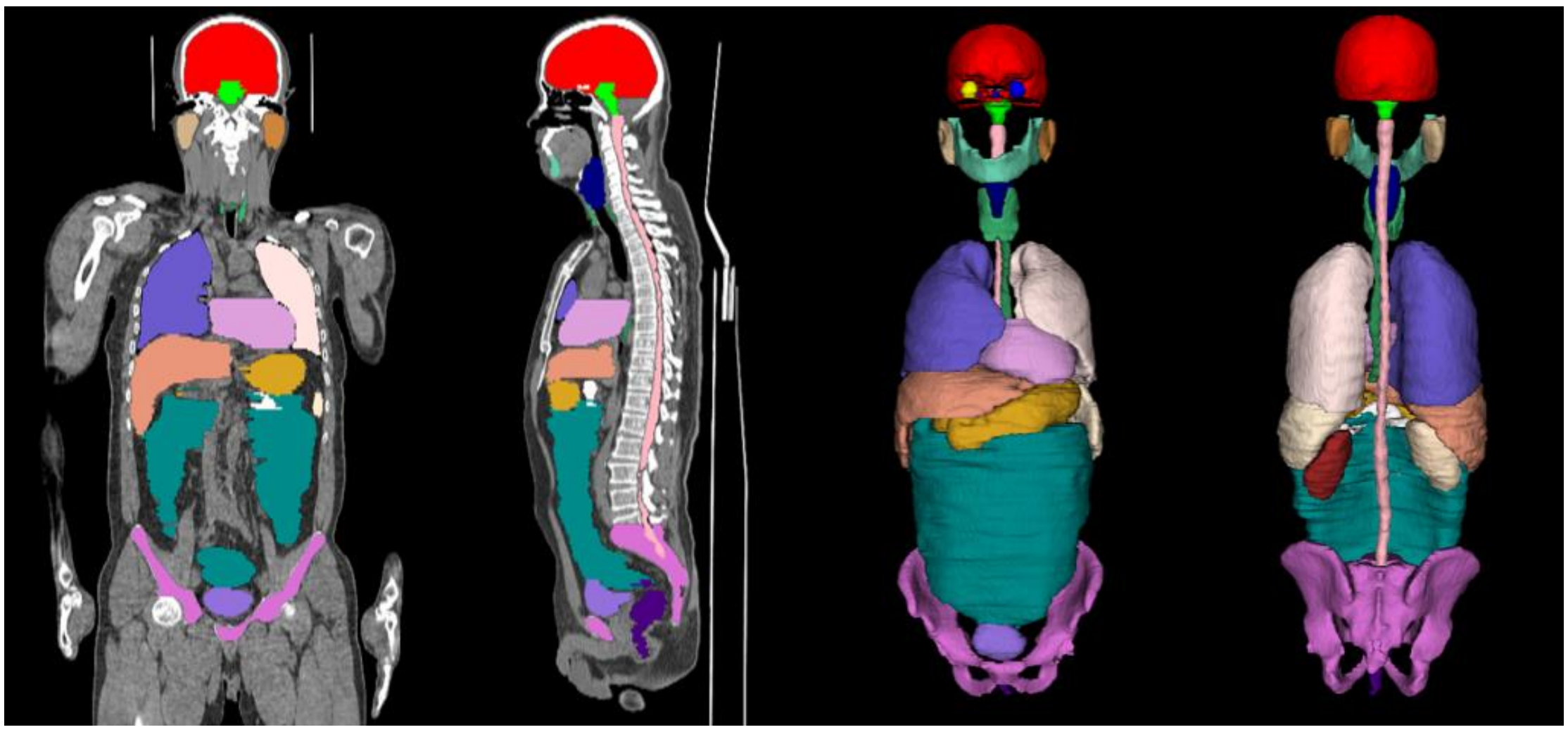

Fig. 4. Visual display of organ autosegmentation results for the $^{18}$F-FDG PET imaging patients

Fig. 5 compares the average dose rate for 27 organs of this patient for GATE and ARCHER-NM. It can be seen that the high-dose-rate areas are mainly in the head —— including the brain, brainstem, optic chiasm, optic nerve, and pituitary —— followed by the bladder and liver. The average dose rate and corresponding statistical uncertainty of these 27 organs are summarized for GATE and ARCHER-NM in Table 1. The relative percent differences are reported, taking GATE results as reference. The dose statistical uncertainties of these 27 organs range from 1.09% to 3.26% for GATE and 1.07% to 3.85% for ARCHER-NM. The relative percent differences for these 27 organs range from -0.58% to 4.11%. There is a very small dose rate difference (less than 1%) in some organs, such as the brain, brain stem, eyeball, and optical nerve. The largest dose rate difference occurs in the stomach. The average absolute value of relative percent differences for 27 organs is 2.19%. Considering the statistical uncertainty, results from ARCHER-NM agree excellently with those from the GATE.

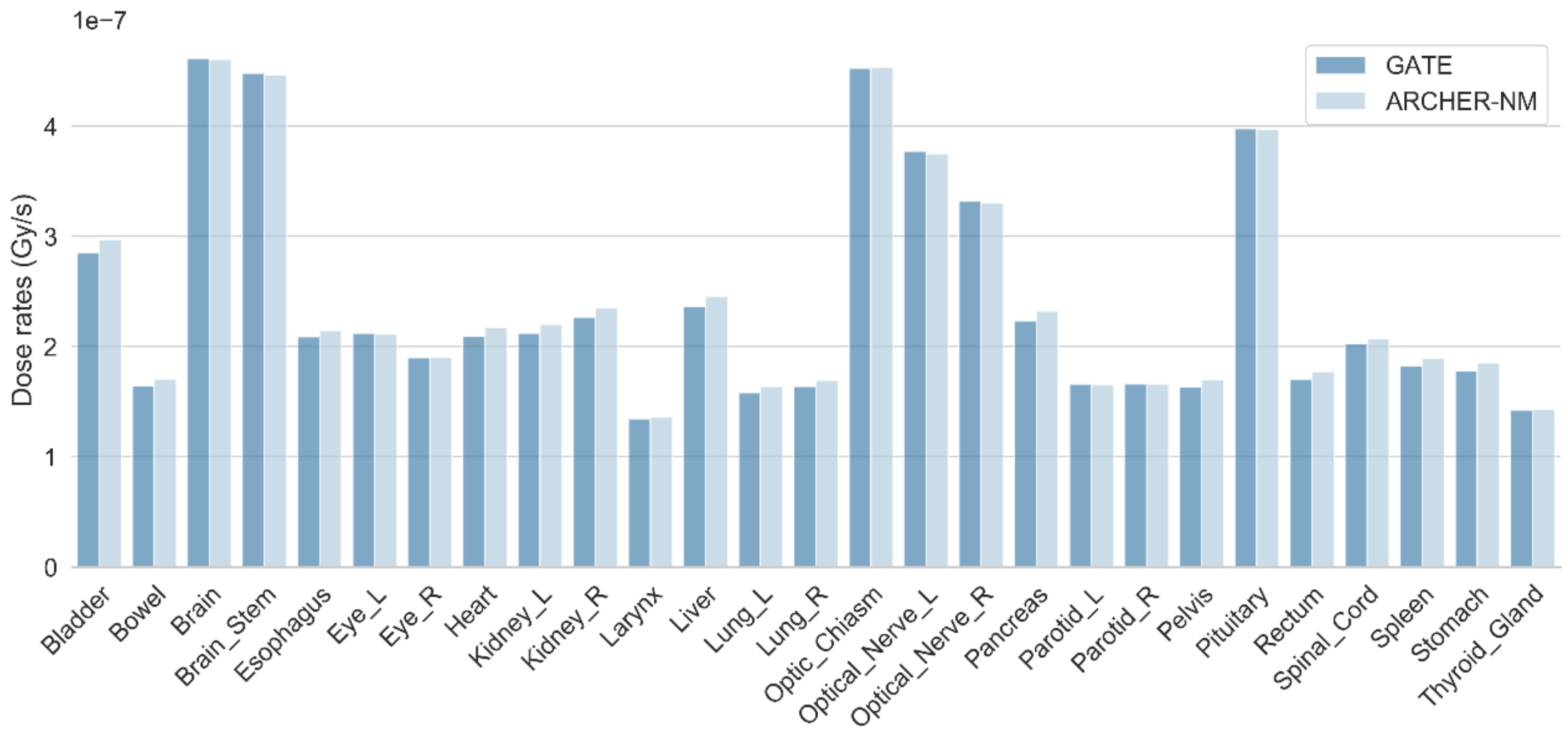

Fig. 5. Comparison of organ dose rate results showing excellent agreement between GATE and ARCHER-NM for $^{18}$F-FDG PET imaging patients.

Table 1. Comparison of the average dose rate and statistical uncertainty (for a total of MC histories of $1 \times 10^9$ for both GATE and ARCHER-NM) in 27 organs calculated by GATE and ARCHER-NM for a patient injected with $^{18}$F-FDG. The percent differences are reported taking GATE results as reference.

| Organs | GATE | | ARCHER-NM | | Relative difference (%) |
|---|---|---|---|---|---|
| | dose rate (Gy/s) | Statistical uncertainty (%) | dose rate (Gy/s) | Statistical uncertainty (%) | |
| Bladder | 2.853E-07 | 1.40 | 2.967E-07 | 3.67 | 4.00 |
| Bowel | 1.641E-07 | 1.82 | 1.706E-07 | 2.83 | 3.96 |
| Brain | 4.612E-07 | 1.10 | 4.605E-07 | 1.11 | -0.15 |
| Brain_Stem | 4.481E-07 | 1.09 | 4.465E-07 | 1.12 | -0.36 |
| Esophagus | 2.090E-07 | 1.62 | 2.145E-07 | 2.52 | 2.63 |
| Eye_Left | 2.119E-07 | 1.62 | 2.111E-07 | 1.59 | -0.38 |
| Eye_Right | 1.899E-07 | 1.69 | 1.902E-07 | 1.72 | 0.16 |
| Heart | 2.094E-07 | 1.57 | 2.170E-07 | 2.76 | 3.63 |
| Kidney_Left | 2.118E-07 | 1.55 | 2.201E-07 | 2.78 | 3.92 |
| Kidney_Right | 2.264E-07 | 1.50 | 2.349E-07 | 2.67 | 3.75 |
| Larynx | 1.343E-07 | 2.05 | 1.358E-07 | 2.08 | 1.12 |
| Liver | 2.362E-07 | 1.48 | 2.455E-07 | 2.73 | 3.94 |
| Lung_Left | 1.584E-07 | 3.23 | 1.637E-07 | 3.81 | 3.35 |
| Lung_Right | 1.640E-07 | 3.26 | 1.694E-07 | 3.85 | 3.29 |
| Optic_Chiasm | 4.527E-07 | 1.09 | 4.534E-07 | 1.07 | 0.15 |
| Optical_Nerve_Left | 3.772E-07 | 1.23 | 3.750E-07 | 1.30 | -0.58 |
| Optical_Nerve_Right | 3.318E-07 | 1.32 | 3.303E-07 | 1.19 | -0.45 |
| Pancreas | 2.233E-07 | 1.48 | 2.323E-07 | 2.61 | 4.03 |
| Parotid_Left | 1.659E-07 | 1.81 | 1.652E-07 | 1.80 | -0.42 |
| Parotid_Righ | 1.663E-07 | 1.82 | 1.659E-07 | 1.87 | -0.24 |
| Pelvis | 1.632E-07 | 1.63 | 1.698E-07 | 2.96 | 4.04 |
| Pituitary | 3.977E-07 | 1.17 | 3.968E-07 | 1.16 | -0.23 |
| Rectum | 1.702E-07 | 1.75 | 1.769E-07 | 3.13 | 3.94 |
| Spinal_Cord | 2.022E-07 | 1.59 | 2.069E-07 | 2.45 | 2.32 |
| Spleen | 1.823E-07 | 1.66 | 1.891E-07 | 2.65 | 3.73 |
| Stomach | 1.778E-07 | 1.73 | 1.851E-07 | 2.54 | 4.11 |
| Thyroid_Gland | 1.426E-07 | 1.88 | 1.429E-07 | 1.92 | 0.21 |

## 4. Discussion

At present, there have been some studies that involve the calculations of organ dose rate for PET imaging patient. For example, Pistone et al. (Pistone *et al.*, 2020) performed the dose rate evaluation using GATE simulations for a case of PET/CT diagnostic exam conducted with $^{18}F$-choline radiopharmaceutical. Lee al et. (Lee *et al.*, 2019) proposed a dose rate estimation method using deep convolutional neural network for personalized internal dosimetry involving PET/CT data set of $^{68}Ga$-NOTA-RGD. The similarities in these studies are that PET images were converted to the distribution maps of the radioactive source for MC simulations. Since the PET images are acquired at a specific time, the result is the dose rate instead of dose. Comparing with the organ dose rate reported in these studies, the results in our study are on the same order of magnitude. In addition, ARCHER-NM needs only 0.5 minutes to achieve acceptable statistical accuracy. Such a speed even exceeds the speed of the deep learning-base method proposed by Lee al el (Lee *et al.*, 2019). It suggests that the GPU-accelerated MC code such as ARCHER-NM has an enormous advantage than those in the literature in terms of both accuracy and calculation speed.

The limitation of this study is that ARCHER-NM provides only dose rate information as the calculations are based on PET images acquired at a certain time. To obtain the total organ dose information for internally deposited radionuclides, we would need multiple PET images at different acquisition time points. Alternatively, of course, it is feasible to use biokinetic models to calculate the total radioactivity distribution (Mattsson *et al.*, 2015). Furthermore, although only the radionuclide $^{18}F$ is simulated in this study, ARCHER-NM can be readily extended to more radionuclides, such as $^{177}Lu$ and $^{131}I$, that emit gamma and beta radiation. Nevertheless, this study shows that ARCHER-NM as a dose engine is ready to be integrated with patient biokinetic information as well as additional radionuclides for applications such as the planning of the radionuclide therapy. In addition to the rapid calculation of the internal radiation dose maps, organ autosegmentation is also very important for organ dose assessment. With the development of artificial intelligence, the accuracy and efficiency of organ segmentations will be further improved. Therefore, combined with the organ autosegmentation technique, ARCHER-NM will have the ability to achieve rapid organ dose assessment for NM in the future.

## 5. Conclusion

In this study, a GPU-accelerated MC dose calculation code dedicated to nuclear medicine internal dosimetry is developed and validated. We first validate the code by calculating the dose distribution in a water phantom for a 0.511-MeV point photon source. Then, the validated code is applied to the calculations of dose rate distributions for an $^{18}F$-FDG PET imaging patient. The results of both cases show that there is an excellent agreement between ARCHER-NM and the widely used MC code GATE. In terms of calculation speed, ARCHER-NM presents an enormous advantage compared with GATE. GATE would need more than 6 hours to achieve acceptable statistical accuracy for the absorbed dose of $^{18}F$-FDG PET imaging patients and this level of computing efficiency is clearly unacceptable for clinical practice. In comparison, ARCHER-NM dose calculation only needs 0.5 minutes. Together with the autosegmentation tool which takes about 10 minutes, accurate and fast patient-specific organ dose assessment is feasible for routine PET/CT imaging and radionuclide therapy procedures.

**Acknowledgements**

This work was supported in part by University of Science and Technology of China (USTC) grants on "New Medicine Team Project: The ROADMAP Medical Physics Platform" and "Med-X Medical Physics and Biomedical Engineering Interdisciplinary Subjects" Strategic Priority Research Program (No. XDB39040600), and in part by Natural Science Foundation of Anhui Province, China (No. 1908085MA27).